\documentclass[a4paper]{article}

\usepackage{amsmath,amsfonts,epsfig}

\newcommand{\rf}[1]{(\ref{#1})}
\newcommand{\Rl}{\mathbb{R}}
\newcommand{\ebar}{\bar{e}}
\newcommand{\nbar}{\bar{n}}
\newcommand{\Trev}{\tilde{T}}
\newcommand{\dt}{\! \cdot \!}
\newcommand{\wdg}{\! \wedge \!}

\newcommand{\del}{\delta}
\newcommand{\eps}{\epsilon}

\newcommand{\lam}{\lambda}

\newcommand{\nn}{\nonumber}
\newcommand{\half}{{\textstyle \frac{1}{2}}}
\newcommand{\Rrev}{\tilde{R}}
\newcommand{\et}[1]{e^{\mbox{\small $#1$}}}
\newcommand{\qrt}{{\textstyle \frac{1}{4}}}
\newcommand{\la}{\langle}
\newcommand{\ra}{\rangle}
\newcommand{\clg}{{\mathcal{G}}}

\begin{document}

\begin{center}

{\bf\Large Conformal geometry, Euclidean space \\ and geometric
algebra}   

\vspace{0.4cm}

Chris Doran\footnote{e-mail: \texttt{c.doran@mrao.cam.ac.uk}}, Anthony
Lasenby\footnote{e-mail: \texttt{a.n.lasenby@mrao.cam.ac.uk}} and 
Joan Lasenby\footnote{e-mail: \texttt{jl@eng.cam.ac.uk}}

\vspace{0.4cm}

Cambridge University, UK

\vspace{0.4cm}

\begin{abstract}
Projective geometry provides the preferred framework for most
implementations of Euclidean space in graphics applications.
Translations and rotations are both linear transformations in
projective geometry, which helps when it comes to programming
complicated geometrical operations.  But there is a fundamental
weakness in this approach --- the Euclidean distance between points is
not handled in a straightforward manner.  Here we discuss a
solution to this problem, based on conformal geometry.  The language
of geometric algebra is best suited to exploiting this geometry, as it
handles the interior and exterior products in a single, unified
framework.  A number of applications are discussed, including a
compact formula for reflecting a line off a general spherical surface.
\end{abstract}

\vspace{0.4cm}

Keywords: geometric algebra, Clifford algebra, conformal geometry, projective
geometry, homogeneous coordinates, sphere geometry, stereographic
projection 

\end{center}

\section{Introduction}

In computer graphics programming the standard framework for modeling
points in space is via a projective representation.  So, for handling
problems in three-dimensional geometry, points in Euclidean space $x$
are represented projectively as rays or vectors in a four-dimensional
space,
\begin{equation}
X = x + e_4.
\end{equation}
The additional vector $e_4$ is orthogonal to $x$, $e_4 \dt x = 0$, and
is normalised to 1, $(e_4)^2=1$.  From the definition of $X$ it is
apparent that $e_4$ is the projective representation of the origin in
Euclidean space.  The projective representation is
\textit{homogeneous}, so both $X$ and $\lam X$ represent the same
point.  Projective space is also not a linear space, as the zero
vector is excluded.  Given a vector $A$ in projective space, the
Euclidean point $a$ is then recovered from
\begin{equation}
a = \frac{A - A \dt e_4 \, e_4}{A \dt e_4}.
\end{equation}
The components of $A$ define a set of homogeneous coordinates for the
position $a$.

The advantage of the projective framework is that the group of
Euclidean transformations (translations, reflections and rotations) is
represented by a set of linear transformations of projective vectors.
For example, the Euclidean translation $x \mapsto x +a$ is described
by the matrix transformation 
\begin{equation}
\begin{pmatrix}
1 & 0 & 0 & a_1 \\
0 & 1 & 0 & a_2 \\
0 & 0 & 1 & a_3 \\
0 & 0 & 0 & 1 
\end{pmatrix}
\begin{pmatrix}
x_1 \\ x_2 \\ x_3 \\ 1
\end{pmatrix}
= 
\begin{pmatrix}
x_1 +a_1 \\ x_2 + a_2 \\ x_3 +a_3 \\ 1
\end{pmatrix}.
\end{equation}
This linearisation of a translation ensures that compounding a
sequence of translations and rotations is a straightforward exercise
in projective geometry.  All one requires for applications is a fast
engine for multiplying together $4\times 4$ matrices.

The main operation in projective geometry is the \textit{exterior
product}, originally introduced by Grassmann in the nineteenth
century~\cite{gra-1862,ste86}.  This product is denoted with the wedge
symbol $\wedge$.  The outer product of vectors is associative and
totally antisymmetric.  So, for example, the outer product of two
vectors $A$ and $B$ is the object $A \wdg B$, which is a rank-2
antisymmetric tensor or \textit{bivector}.  The components of $A \wdg
B$ are 
\begin{equation}
(A \wdg B)_{ij} = A_i B_j - A_j B_i.
\end{equation}
The exterior product defines the \textit{join} operation in projective
geometry, so the outer product of two points defines the line between
them, and the outer product of three points defines a plane.  In this
scheme a line in three dimensions is then described by the 6
components of a bivector.  These are the Pl\"{u}cker coordinates of a
line.  The associativity and antisymmetry of the outer product ensure
that
\begin{equation}
(A \wdg B) \wdg (A \wdg B) = A \wdg B \wdg A \wdg B = 0,
\end{equation}
which imposes a single quadratic condition on the coordinates of a
line.  This is the Pl\"{u}cker condition.

The ability to handle straight lines and planes in a systematic manner
is essential to practically all graphics applications, which explains
the popularity of the projective framework.  But there is one crucial
concept which is missing.  This is the Euclidean \textit{distance}
between points.  Distance is a fundamental concept in the Euclidean
world which we inhabit and are usually interested in modeling.  But
distance cannot be handled elegantly in the projective framework, as
projective geometry is non-metrical.  Any form of distance measure
must be introduced via some additional structure.  One way to proceed
is to return to the Euclidean points and calculate the distance
between these directly.  Mathematically this operation is distinct
from all others performed in projective geometry, as it does not
involve the exterior product (or duality).  Alternatively, one can
follow the route of classical planar projective geometry and define
the additional metric structure through the introduction of the
\textit{absolute conic}~\cite{cdy-man}.  But this structure requires
that all coordinates are complexified, which is hardly suitable for
real graphics applications.  In addition, the generalisation of the
absolute conic to three-dimensional geometry is awkward.

There is little new in these observations.  Grassmann himself was
dissatisfied with an algebra based on the exterior product alone, and
sought an algebra of points where distances are encoded in a natural
manner.  The solution is provided by the \textit{conformal model} of
Euclidean geometry, originally introduced by M\"{o}bius in his study
of the geometry of spheres.  The essential new feature of this space
is that it has mixed signature, so the inner product is not positive
definite.  In the nineteenth century, when these developments were
initiated, mixed signature spaces were a highly original and somewhat
abstract concept.  Today, however, physicists and mathematicians
routinely study such spaces in the guise of special relativity, and
there are no formal difficulties when computing with vectors in these
spaces.  As a route to understanding the conformal representation of
points in Euclidean geometry we start with a description of the
\textit{stereographic projection}.  This map provides a means of
representing points as null vectors in a space of two dimensions
higher than the Euclidean base space.  This is the conformal
representation.  The inner product of points in this space recovers
the Euclidean distance, providing precisely the framework we desire.
The outer product extends the range of geometric primitives from
projective geometry to include circles and spheres, which has many
applications.

The conformal model of Euclidean geometry makes heavy use of both the
interior and exterior products.  As such, it is best developed in the
language of \textit{geometric algebra} --- a universal language for
geometry based on the mathematics of \textit{Clifford
algebra}~\cite{hes-gc,DL-gap,DL-course}.  This is described in
section~\ref{Sga}.  The power of the geometric algebra development
becomes apparent when we discuss the group of conformal
transformations, which include Euclidean transformations as a
subgroup.  As in the projective case, all Euclidean transformations
are linear transformations in the conformal framework.  Furthermore,
these transformations are all \textit{orthogonal}, and can be built up
from primitive reflections.

The join operation in conformal space generalises the join of
projective geometry.  Three points now define a line, which is the
circle connecting the points.  If this circle passes through the point
at infinity it is a straight line.  Similarly, four points define a
sphere, which reduces to a plane when its radius is infinite.  These
new geometric primitives provide a range of intersection and
reflection operations which dramatically extend the available
constructions in projective geometry.  For example, reflecting a line
in a sphere is encoded in a simple expression involving a pair of
elements in geometric algebra.  Working in this manner one can write
computer code for complicated geometrical operations which is robust,
elegant and highly compact.  This has many potential applications for
the graphics industry.

\section{Stereographic projection and conformal space}

\begin{figure}
\begin{center}
\includegraphics[width=9cm]{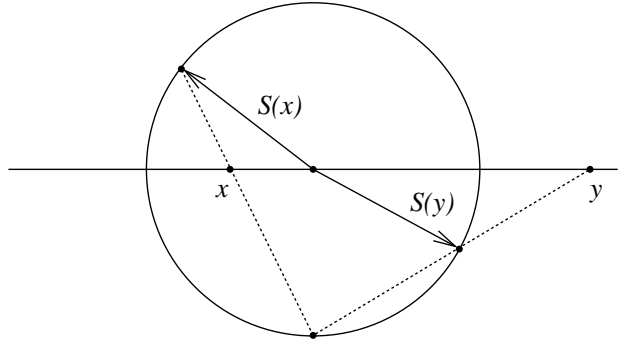}
\end{center}
\caption[A stereographic projection]{\textit{A stereographic
projection}.  The space $\Rl^n$ is mapped to the unit sphere
$S^n$.  Given a point in $\Rl^n$ we form the line through this
point and the south pole of the sphere.  The point where this line
intersects the sphere defines the image of the projection.}
\label{Fstereog}
\end{figure}

The stereographic projection provides a straightforward route to the
principle construction of the conformal model --- the representation
of a point as a null vector in conformal space.  The stereographic
projection maps points in the Euclidean space $\Rl^n$ to points
on the unit sphere $S^n$, as illustrated in figure~\ref{Fstereog}.
Suppose that the initial point is given by $x \in \Rl^n$, and
we write
\begin{equation}
x = r \hat{x},
\end{equation}
where $r$ is the magnitude of the vector $x$, $r^2=x^2$.  The
corresponding point on the sphere is
\begin{equation}
S(x) = \cos\!\theta \, \hat{x} - \sin\!\theta \, e,
\end{equation}
where $e$ is the unit vector perpendicular to the plane defining the
south pole of the sphere $S^n$.  The angle $\theta$, $-\pi/2 \leq
\theta \leq \pi/2$ is related to the distance $r$ by
\begin{equation}
r = \frac{\cos\!\theta}{1+\sin\!\theta},
\end{equation}
which inverts to give
\begin{equation}
\cos\theta =  \frac{2r}{1+r^2}, \quad 
\sin\theta = \frac{1-r^2}{1+r^2}.
\end{equation}
The stereographic projection $S$ maps $\Rl^n$ into $S^n-{P}$, where
$P$ is the south pole of $S^n$.  We complete the space $S^n$ by
letting the south pole represent the point at infinity.  We therefore
expect that, under Euclidean transformations of $\Rl^n$, the point at
infinity should remain invariant.

We now have a representation of points in $\Rl^n$ with unit vectors in
the space $\Rl^{n+1}$.  But the constraint that the vector has unit
magnitude means that this representation is not homogeneous.  A
homogeneous representation of geometric objects is critical to the
power of projective geometry, as it enables us to write the equation
of the line through $A$ and $B$ as
\begin{equation}
A \wdg B \wdg X = 0.
\end{equation}
Clearly, if $X$ satisfies this equation, then so to does $\lam X$.  To
achieve a homogeneous representation we introduce a 
further vector, $\ebar$, which has \textit{negative
signature},
\begin{equation} 
\ebar^2= -1.
\end{equation}
We also assume that $\ebar$ is orthogonal to $x$ and $e$.  We can now
replace the unit vector $S(x)$ with the vector $X$, where
\begin{equation}
X =  S(x) + \ebar = \frac{2x}{1+x^2} - \frac{1-x^2}{1+x^2}  e + \ebar.
\label{EX1}
\end{equation}
The vector $X$ satisfies 
\begin{equation}
X \dt X =0
\end{equation}
so is \textit{null}.  This equation is homogeneous, so we can now move
to a homogeneous encoding of points and let $X$ and $\lam X$ represent
the \textit{same} point in $\Rl^n$.  Multiplying the vector in
equation~\rf{EX1} by $(1+x^2)$ we establish the conformal
representation
\begin{equation} 
F(x) = X = 2x - (1-x^2) e + (1+x^2) \ebar.
\end{equation}
The vectors $e$ and $\ebar$ extend the Euclidean space $\Rl^n$ to a
space with two extra dimensions and signature $(n+1,1)$.  It is
generally more convenient to work with a null basis for the extra
dimensions, so we define
\begin{equation}
n  = e + \ebar  \qquad  \nbar  = e - \ebar.
\end{equation}
These vectors satisfy
\begin{equation}
n^2 = \nbar ^2 = 0, \qquad n \dt \nbar = 2.
\end{equation}
The vector $X$ is now 
\begin{equation}
F(x) = X = 2x + x^2 n - \nbar,
\label{Eptmap}
\end{equation}
which defines our standard representation of points as vectors in
conformal space.  Given a general, unnormalised null vector in
conformal space, the standard form of equation~\rf{Eptmap} is
recovered by setting
\begin{equation}
X \mapsto -2 \frac{X}{X \dt n}.
\end{equation}
This map makes it clear that the null vector $n$ now represents the
point at infinity.  In general we will not assume that our points are
normalised, so the components of $X$ are homogeneous coordinates for
the point $x$.  The step of normalising the representation is only
performed if the actual Euclidean point is required.

Given two null vectors $X$ and $Y$, in the form of
equation~\rf{Eptmap}, their inner product is
\begin{align}
X \dt Y &= \bigl( x^2n + 2x -\nbar  \bigr) \dt \bigl( y^2n + 2y -
\nbar \bigr) \nn \\
&=-2x^2 -2y^2 +4x\dt y \nn \\
&= -2(x-y)^2.  
\label{Econfinner}
\end{align}
This is the key result which justifies the conformal model approach to
Euclidean geometry.  The inner product in conformal space encodes the
\textit{distance} between points in Euclidean space.  This is why
points are represented with null vectors --- the distance between a
point and itself is zero.  Since equation~\rf{Econfinner} was
appropriate for normalised points, the general expression relating
Euclidean distance to the conformal inner product is

\begin{equation}
|x-y|^2 = -2 \frac{X \dt Y}{X \dt n \, Y \dt n}.
\label{Eucdist}
\end{equation}
This is manifestly homogeneous in $X$ and $Y$.  This formula returns
the dimensionless distance.  To introduce dimensions one requires a
fundamental length scale $\lam$, so that $x$ is the dimensionless
representation of the position vector $\lam x$.  Appropriate factors
of $\lam$ can then be inserted when required.

An orthogonal transformation in conformal space will ensure that a
null vector remains null.  Such a transformation therefore maps points
to points in Euclidean space. This defines the full conformal group of
Euclidean space, which is isomorphic to the group $\mbox{SO}(n+1,1)$.
Conformal transformations leave angles invariant, but can map straight
lines into circles.  The Euclidean group is the subgroup of the
conformal group which leaves the Euclidean distance invariant.  These
transformations include translations and rotations, which are
therefore linear, orthogonal transformations in conformal space.  The
key to developing simple representations of conformal transformations
is geometric algebra, which we now describe.

\section{Geometric algebra}
\label{Sga}

The language of geometric algebra can be thought of as Clifford
algebra with added geometric content.  The details are described in
greater detail elsewhere \cite{hes-gc,DL-gap,DL-course}, and here we
just provide a brief introduction.  A geometric algebra is constructed
on a vector space with a given inner product.  The \textit{geometric}
product of two vectors $a$ and $b$ is defined to be associative and
distributive over addition, with the additional rule that the
(geometric) square of any vector is a scalar,
\begin{equation}
aa = a^2 \in \Rl.
\end{equation}
If we write
\begin{equation}
ab + ba = (a+b)^2 - a^2 - b^2
\end{equation}
we see that the symmetric part of the geometric product of any two
vectors is also a scalar.  This defines the inner product, and we
write
\begin{equation}
a \dt b = \half(ab+ba).
\end{equation}
The geometric product of two vectors can now be written
\begin{equation}
ab = a \dt b + a \wdg b, 
\end{equation}
where the exterior product is the antisymmetric combination
\begin{equation}
a \wdg b = \half(ab-ba).
\end{equation}
Under the geometric product, orthogonal vectors anticommute and
parallel vectors commute.  The product therefore encodes the basic
geometric relationships between vectors.  The totally antisymmetrised
sum of geometric products of vectors defines the exterior product in
the algebra.

Once one knows how to multiply together vectors it is a
straightforward exercise to construct the entire geometric algebra of
a vector space.  General elements of this algebra are called
\textit{multivectors}, and they too can be multiplied via the
geometric product.  The two algebras which concern us in this paper
are the algebras of conformal vectors for the Euclidean plane and
three-dimensional space \cite{hes-som1,hes-som2,LL-surf}.  For the
Euclidean plane, let $\{e_1,e_2\}$ denote an orthonormal basis set,
and write $e_0=\ebar$, $e_3=e$.  The full basis set is then $\{e_i\},
i=0\ldots 3$.  These generators satisfy
\begin{equation}
e_i e_j + e_j e_i = 2 \eta_{ij}, \qquad i,j=0\ldots 3,
\end{equation}
where
\begin{equation}
\eta_{ij} = \mbox{diag}(-1,1,1,1).
\end{equation}
The algebra generated by these vectors consists of the set
\begin{equation*} 
\begin{array}{ccccc}
1 & \{e_i \} & \{e_i \wdg e_j \}  & \{I e_i \} & I \\
\mbox{{1 scalar}} & 
\mbox{{4 vectors}} & 
\mbox{{6 bivectors}} & 
\mbox{{4 trivectors}} & 
\mbox{{1 pseudoscalar.}}
\end{array}
\end{equation*} 
Scalars are assigned grade zero, vectors grade one, and so on.  The
highest grade term in the algebra is the pseudoscalar $I$, 
\begin{equation}
I = e_0 e_1 e_2 e_3 = e_1 e_2 \ebar e.
\label{Eps2}
\end{equation}
This satisfies
\begin{equation}
I^2 = e_1 e_2 \ebar e e_1 e_2 \ebar e = - e_1 e_2 \ebar e_1 e_2 \ebar
= e_1 e_2 e_1 e_2 = - e_1 e_1 = -1.
\end{equation}
The steps in this calculation simply involve counting the number of
sign changes as a vector is anticommuted past orthogonal vectors.
This is essentially how all products in geometric algebra are
calculated, and it is easily incorporated into any programming
language.  In even dimensions, such as the case here, the pseudoscalar
anticommutes with vectors,
\begin{equation}
I a = -a I, \quad \mbox{even dimensions}.
\end{equation}

For the case of the conformal algebra of Euclidean three-dimensional
space $\Rl^3$, we can define a basis as $\{e_i\}, i=0\ldots 4$, with
$e_0=\ebar$, $e_4=e$, and $e_i, i=1\ldots 3$ a basis set for
three-dimensional space.  The algebra generated by these vectors has
32 terms, and is spanned by
\begin{equation*} 
\begin{array}{rcccccc}
& 1 & \{e_i \} & \{e_i \wdg e_j \}  &  \{ e_i \wdg e_j \wdg e_k \} & \{I
e_i \} & I \\ 
\mbox{grade} & 0 & 1 & 2 & 3 & 4 & 5\\
\mbox{dimension} & 1  & 5  & 10 & 10  & 5 & 1.
\end{array}
\end{equation*} 
The dimensions of each subspace are given by the binomial
coefficients.  Each subspace has a simple geometric interpretation in
conformal geometry.  The pseudoscalar for five-dimensional space is
again denoted by $I$, and this time is defined by
\begin{equation}
I = e_0 e_1 e_2 e_3 e_4 = e_1 e_2 e_3 e \ebar.
\label{Eps3}
\end{equation}
In five-dimensional space the pseudoscalar commutes with all elements
in the algebra.  The $(4,1)$ signature of the space implies that the
pseudoscalar satisfies
\begin{equation}
I^2 = -1.
\end{equation}
So, algebraically, $I$ has the properties of a unit imaginary, though
in geometric algebra it plays a definite geometric role.  In a general
geometric algebra, multiplication by the pseudoscalar performs the
duality transformation familiar in projective geometry.

\section{Reflections and rotations in geometric algebra}

Suppose that the vectors $a$ and $m$ represent two lines from a common
origin in Euclidean space, and we wish to reflect the vector $a$ in
the hyperplane perpendicular to $m$.  If we assume that $m$ is
normalised to $m^2=1$, the result of this reflection is
\begin{equation}
a \mapsto a' = a - 2 (a \dt m) m.
\end{equation}
This is the standard expression one would write down without access to
the geometric product.  But with geometric algebra at our disposal we
can expand $a'$ into
\begin{equation}
a' = a - (am+ma)m = a - am^2 - mam = -mam.
\end{equation}
The advantage of this representation of a reflection is that we can
easily chain together reflections into a series of geometric
products.  So two reflections, one in $m$ followed by one in $l$,
produce the transformation 
\begin{equation}
a \mapsto lm a ml.
\end{equation}
But two reflections generate a rotation, so a rotation in geometric
algebra can be written in the simple form
\begin{equation}
a \mapsto R a \Rrev
\label{Erotn}
\end{equation}
where
\begin{equation} 
R = lm, \qquad \Rrev = ml.
\end{equation}
The tilde denotes the operation of \textit{reversion}, which reverses
the order of vectors in any series of geometric products.  Given a
general multivector $A$ we can decompose it into terms of a unique
grade by writing 
\begin{equation}
A = A_0 + A_1 + A_2 + \cdots,
\end{equation}
where $A_r$ denotes the grade-$r$ part of $A$.  The effect of the
reverse on $A$ is then
\begin{equation}
\tilde{A} = A_0 + A_1 - A_2 - A_3 + A_4 + \cdots.
\end{equation}
The geometric product of an even number of positive norm unit vectors
is called a rotor.  These satisfy $R\Rrev=1$ and generate rotations.
A rotor can be written as
\begin{equation}
R = \pm \exp(B/2)
\end{equation}
where $B$ is a bivector.  The space of bivectors is therefore the
space of generators of rotations.  These define a \textit{Lie
algebra}, with the rotors themselves defining a \textit{Lie group}.
The action of this group on vectors is defined by equation~\rf{Erotn},
so both $R$ and $-R$ define the same rotation.

\section{Euclidean and conformal transformations}

Transformations of conformal vectors which leave the product of
equation~\rf{Eucdist} invariant correspond to the group of Euclidean
transformations in $\Rl^n$.  To simplify the notation, we let
$\clg(p,q)$ denote the geometric algebra of a vector space with
signature $(p,q)$.  The Euclidean spaces of interest to us therefore
have the algebras $\clg(2,0)$ and $\clg(3,0)$ associated with them.
The corresponding conformal algebras are $\clg(3,1)$ and $\clg(4,1)$
respectively.  Each Euclidean algebra is a subalgebra of the
associated conformal algebra.

The operations which mainly interest us here are translations and
rotations.  The fact that translations can be treated as orthogonal
transformations is a novel feature of conformal geometry.  This is
possible because the underlying orthogonal group is non-compact, and
so contains null generators.  To see how this allows us to describe a
translation, consider the rotor
\begin{equation}
R=T_a = \et{na/2}
\end{equation}
where $a$ is a vector in the Euclidean space, so that $a \dt n=0$.
The bivector generator satisfies
\begin{equation}
(na)^2 = -anna = 0,
\end{equation}
so is null.  The Taylor series for $T_a$ therefore terminates after
two terms, leaving
\begin{equation}
T_a  = 1 + \frac{na}{2} .
\end{equation}
The rotor $T_a$ transforms the null vectors $n$ and $\nbar$ into
\begin{equation}
T_a n\Trev_a  
= n + \half nan + \half nan + \qrt nanan =  n
\end{equation}
and
\begin{equation} 
T_a \nbar \Trev_a = \nbar  - 2a - a^2n.
\end{equation}
As expected, the point at infinity remains at infinity, whereas the
origin is transformed to the vector $a$.  
Acting on a vector $x\in\clg(n,0)$ we similarly obtain
\begin{equation} 
T_a x\Trev_a =  x + n(a\dt x).
\end{equation}
Combining these results we find that
\begin{align}
T_a F(x) \Trev_a 
&= x^2n + 2(x + a\dt x \, n) - (\nbar  - 2a - a^2n)   \nn \\
&= (x+a)^2n + 2(x+a) - \nbar  \nn \\
&= F(x+a),
\end{align}
which performs the conformal version of the translation $x \mapsto
x+a$.  Translations are handled as rotations in conformal space, and
the rotor group provides a double-cover representation of a
translation.  The identity
\begin{equation}
\Trev_a = T_{-a}
\end{equation}
ensures that the inverse transformation in conformal space corresponds
to a translation in the opposite direction, as required.

Similarly, as discussed above, a rotation in the origin in $\Rl^n$ is
performed by $x \mapsto x' =Rx\Rrev$, where $R$ is a rotor in
$\clg(n,0)$.  The conformal vector representing the transformed point is
\begin{align}
F(x') &= (x')^2 n + 2 R x \Rrev - \nbar \nn \\
&= R(x^2 n + 2x -\nbar)\Rrev = R F(x) \Rrev.
\end{align}
This holds because $R$ is an even element in $\clg(n,0)$, so must commute
with both $n$ and $\nbar$.  Rotations about the origin therefore take
the same form in either space.  But suppose instead that we wish to
rotate about the point $a \in \Rl^n$.  This can be achieved by
translating $a$ to the origin, rotating, and then translating forward
again.  In terms of $X=F(x)$ the result is
\begin{equation}
X \mapsto T_a R T_{-a} X \tilde{T}_{-a} \Rrev \Trev_a = R' X \Rrev.
\end{equation}
The rotation is now controlled by the rotor $R'$, where
\begin{equation}
R' = T_a R \Trev_{a} = \left(1+ \frac{na}{2}\right) R \left(1+
\frac{an}{2}\right).
\end{equation}
The conformal model frees us up from treating the origin as a special
point.  Rotations about any point are handled with rotors in the same
manner.  Similar comments apply to reflections, though we will see
shortly that the range of possible reflections is enhanced in the
conformal model.  The Euclidean group is a subgroup of the full
conformal group, which consists of transformations which preserve
angles alone.  This group is described in greater detail
elsewhere~\cite{DL-gap,LL-surf}.  The essential property of a
Euclidean transformation is that the point at infinity is invariant,
so all Euclidean transformations map $n$ to itself.

\section{Geometric primitives in conformal space}
\label{Sprims}

In the conformal model, points in Euclidean space are represented
homogeneously by null vectors in conformal space.  As in projective
geometry, a multivector $L\in\clg(n+1,1)$ encodes a geometric object
in $\Rl^n$ via the equations
\begin{equation}
L \wdg X = 0, \qquad X^2=0.
\label{Ecf1}
\end{equation}
One result we can exploit is that $X^2=0$ is unchanged if $X \mapsto
RX\Rrev$, where $R$ is a rotor in $\clg(n+1,1)$.  So, if a geometric object
is specified by $L$ via equation~\rf{Ecf1}, it follows that
\begin{equation}
R (L \wdg X) \Rrev = (RL\Rrev) \wdg (RX\Rrev) = 0.
\end{equation}
We can therefore transform the object $L$ with a general element of
the full conformal group to obtain a new object.  As well as
translations and rotations, the conformal group includes dilations and
inversions, which map straight lines into circles.  The range of
geometric primitives is therefore extended from the projective case,
which only deals with straight lines.

The first case to consider is a pair of null vectors $A$ and $B$.
Their inner product describes the Euclidean distance between points,
and their outer product defines the bivector
\begin{equation}
G = A \wdg B.
\end{equation}
The bivector $G$ has magnitude
\begin{equation}
G^2 = (AB-A \dt B)(-BA + A \dt B) = (A \dt B)^2,
\end{equation}
which shows that $G$ is \textit{timelike}, in the terminology
of special relativity.  It follows that $G$ contains a pair of null
vectors.  If we look for solutions to the equation 
\begin{equation}
G \wdg X = 0, \qquad X^2=0,
\end{equation}
the only solutions are the two null vectors contained in $G$.  These
are precisely $A$ and $B$, so the bivector encodes the two points
directly.  In the conformal model, no information is lost in forming
the exterior product of two null vectors.  Frequently, bivectors are
obtained as the result of intersection algorithms, such as the
intersection of two circles in a plane.  The sign of the square of the
resulting bivector, $B^2$, defines the number of intersection points
of the circles.  If $B^2 >0$ then $B$ defines two points, if $B^2 = 0$
then $B$ defines a single point, and if $B^2<0$ then $B$ contains no
points.

Given that bivectors now define pairs of points, as opposed to lines,
the obvious question is how do we encode lines?  Suppose we construct
the line through the points $a,b \in \Rl^n$.  A point on the line is
then given by
\begin{equation}
x = \lam a + (1-\lam) b.
\end{equation}
The conformal version of this line is
\begin{align}
F(x) &= \bigl(\lam^2 a^2 + 2 \lam (1-\lam) a \dt b + (1-\lam)^2 b
\bigr) n + 2\lam a + 2(1-\lam)b - \nbar \nn \\
&= \lam A + (1-\lam) B + \half \lam(1-\lam) A \dt B \, n,
\end{align}
and any multiple of this encodes the same point on the line.  It
is clear, then, that a conformal point $X$ is a linear combination of
$A$, $B$ and $n$, subject to the constraint that $X^2=0$.  This is
summarised by
\begin{equation}
(A \wdg B \wdg n) \wdg X = 0, \qquad X^2=0.
\end{equation}
So it is the \textit{trivector} $A \wedge B \wedge n$ which represents
a line in conformal geometry.  This illustrates a general feature of
the conformal model --- geometric objects are represented by
multivectors one grade higher than their projective counterpart.
The extra degree of freedom is absorbed by the constraint that
$X^2=0$.

Now suppose that we form a general trivector $L$ from three null
vectors,
\begin{equation}
L = A_1 \wdg A_2 \wdg A_3.
\end{equation}
This must still encode a conformal line via the equation $L \wdg X =
0$.  In fact, $L$ encodes the \textit{circle} through the points
defined by $A_1$, $A_2$ and $A_3$.  To see why, consider the conformal
model of a plane.  The trivector $L$ therefore maps to a \textit{dual}
vector $l$, where
\begin{equation}
l = IL
\end{equation}
and $I$ is the pseudoscalar defined in equation~\rf{Eps2}.  We see
that
\begin{equation}
l^2 = L^2 = -2 (A_1 \dt A_2) (A_1 \dt A_3) (A_2 \dt A_3)
\end{equation}
so $l$ is a vector with positive square.  Such a vector can always be
written in the form
\begin{equation}
l = \lam(F(c) - \rho^2 n) = \lam(C - \rho^2 n),
\end{equation}
where $C=F(c)$ is the conformal vector for the point $c \in\Rl^n$.
The dual version of the equation $X \wdg L = 0$ is
\begin{equation}
X \dt l = 0,
\end{equation}
which reduces to
\begin{equation}
\frac{X \dt C}{X \dt n} = \rho^2.
\label{Ecirc}
\end{equation}
Since $C=F(c)$ satisfies $C \cdot n = -2$, equation~\rf{Ecirc} states
that the Euclidean distance between $x$ and $c$ is equal to the
constant $\rho$.  This clearly defines a circle in a plane.
Furthermore, the radius of the circle is defined by
\begin{equation}
\rho^2 = \frac{l^2}{(l \dt n)^2} = - \frac{L^2}{(L\wdg n)^2},
\end{equation}
where the minus sign in the final expression is due to the fact that
the 4-vector $L \wedge n$ has negative square.  This equation
demonstrates how the conformal framework allows us to encode
dimensional concepts such as radius while keeping multivectors like
$L$ as homogeneous representations of geometric objects.

The equation for the radius of the circle tells us that the circle has
infinite radius if
\begin{equation}
L \wdg n = 0.
\end{equation}
This is the case of a straight line, and this equation can be
interpreted as saying the line passes through the point at infinity.
So, given three points, the test that they lie on a line is 
\begin{equation}
A \wdg B \wdg C \wdg n = 0 \quad \implies \mbox{$A$, $B$, $C$
collinear}.
\label{Elntst}
\end{equation}
This test has an important advantage over the equivalent test in
projective geometry.  The degree to which the right-hand side differs
from zero directly measures how far the points are from lying on a
common line.  This can resolve a range of problems caused by the
finite numerical precision of most computer algorithms.  Numerical
drift can only affect how near to being straight the line is.  When it
comes to plotting, one can simply decide what tolerance is required
and modify equation~\rf{Elntst} to read
\begin{equation}
- \frac{(L \wdg n)^2}{L^2} < \eps \quad \implies \mbox{sufficiently
  straight},
\end{equation}
where $L=A \wedge B \wedge C$.  A similar idea cannot be applied so
straightforwardly in the projective framework, as there is no intrinsic
measure of being `nearly linearly dependent' for projective vectors.

Next, suppose that $a$ and $b$ define two vectors in $\Rl^n$, both as
rays from the origin, and we wish to find the angle between these.
The conformal representation of the line can be built up from
\begin{equation}
F(a) \wdg F(\lam a) \wdg F(\mu a) \propto a e \ebar = aN,
\end{equation}
where $N=e\ebar$ and $N^2=1$.  Similarly the line in the $b$ direction
is represented by $bN$.  We can therefore write
\begin{equation}
L_1 = a N, \qquad L_2 = b N,
\end{equation}
so that
\begin{equation}
L_1 L_2 = ab.
\end{equation} 
If we let angle brackets $\la M \ra$ denote the scalar part of the
multivector $M$ we see that
\begin{equation}
\frac{\la L_1 L_2 \ra}{|L_1| \, |L_2|} = \frac{a \dt b}{|a| \, |b|} =
\cos \! \theta, 
\end{equation}
where $\theta$ is the angle between the lines.  This is true for lines
through the origin, but the expression in terms of $L_1$ and $L_2$ is
unchanged under the action of a general rotor, so applies to lines
meeting at any point, and to circles as well as straight lines.

Similar considerations apply to circles and planes in three
dimensional space.  Suppose that the points $a,b,c$ define a plane in
$\Rl^n$, so that an arbitrary point in the plane is given by
\begin{equation}  
x = \alpha a + \beta b + \gamma c,  \qquad \alpha + \beta +
\gamma = 1.  
\end{equation} 
The conformal representation of $x$ is
\begin{equation}
   X = \alpha A + \beta B + \gamma C + \delta n 
\end{equation}
where $A=f(a)$ \textit{etc.}, and 
\begin{equation}
\del = \half (\alpha\beta A\dt B + \alpha \gamma A\dt C +
\beta\gamma B\dt C).
\end{equation}
Varying $\alpha$ and $\beta$, together with the freedom to scale $F(x)$,
now produces general null combinations of the vectors $A$, $B$, $C$
and $n$.  The equation for the plane can then be written
\begin{equation}
A \wdg B \wdg C \wdg n \wdg X = 0, \qquad X^2 = 0.
\end{equation}
So, as one now expects, it is 4-vectors which define planes.  If
instead we form the 4-vector
\begin{equation}
S = A_1 \wdg A_2 \wdg A_3 \wdg A_4
\end{equation}
then in general $S \wdg X = 0$ defines a sphere.  To see why, consider
the dual of $S$ ,
\begin{equation}
s = IS,
\end{equation}
where $I$ is now given by equation~\rf{Eps3}.  Again we find that
$s^2>0$, and we can write
\begin{equation} 
s = \lam(F(c) - \rho^2 n) = \lam(C - \rho^2 n).
\end{equation}
The equation $X \wdg S = 0$ is now equivalent to $X \dt s = 0$, which
defines a sphere of radius $\rho$, with centre $C=F(c)$.  The radius
of the sphere is defined by
\begin{equation}
\rho^2 = \frac{s^2}{(s \dt n)^2} =  \frac{S^2}{(S \wdg n)^2}.
\end{equation}
The sphere becomes a flat plane if the radius is infinite, so the test
that four points lie on a common plane is
\begin{equation}
A_1 \wdg A_2 \wdg A_3 \wdg A_4 \wdg n = 0 \quad \implies \mbox{$A_1
\ldots A_4$ coplanar}.
\end{equation}
As with the case of the test of collinearity, this is numerically well
conditioned, as the deviation from zero is directly related to the
curvature of the sphere through the four points.

\section{Reflection and intersection in conformal space}

The conformal model extends the range of geometric primitives beyond
the simple lines and planes of projective geometry.  It also provides
a range of new algorithms for handling intersections and reflections.
First, suppose we wish to find the intersection of two lines in a
plane.  These are defined by the trivectors $L_1$ and $L_2$.  The
intersection, or meet, is defined via its dual by
\begin{equation}
(L_1 \vee L_2)^\ast = L_1^\ast \wdg L_2^\ast.
\end{equation}
The star, $L^\ast$, denotes the dual of a multivector, which in
geometric algebra is formed by multiplication by the multivector
representing the join.  For the case of two distinct lines in a plane,
their join is the plane itself, so the dual is formed by
multiplication by the pseudoscalar $I$ of equation~\rf{Eps2}.  The
intersection of two lines therefore results in the bivector
\begin{equation}
B = (L_1 \vee L_2)^\ast = (IL_1) \dt L_2 = L_1 \dt (I L_2).
\end{equation}
This expression is a bivector as it is the contraction of a vector and
a trivector.  As discussed in section~\ref{Sprims}, a bivector can
encode zero, one or two points, depending on the sign of its square.
Two circles can intersect in two points, for example, if they are
sufficiently close to one another.  Two straight lines will also
intersect in two points, though one of these is at infinity.

In three dimensions a number of further possibilities arise.  The
intersection of a line $L$ and a sphere or plane $P$ also results in a
bivector,
\begin{equation}
L \vee P = L \dt (IP),
\end{equation}
where $I$ is now the grade-5 pseudoscalar in the conformal algebra for
three-dimensional Euclidean space.  The bivector encodes the fact that
a line can intersect a sphere in up to two places.  If $P$ is a flat
plane and $L$ a straight line, then one of the intersection points
will be at infinity.  More complex is the case of two planes or
spheres.  In this case both geometric primitives are 4-vectors, $P_1$
and $P_2$.  Their intersection results in the trivector
\begin{equation}
L = P_1 \vee P_2 = (IP_1) \dt P_2 = P_1 \dt (IP_2).
\end{equation}
This trivector encodes a line.  This one single expression covers a
wide range of situations, since either plane can be flat or spherical.
If both planes are flat their intersection is a straight line with $L
\wdg n = 0$.  If one or both of the planes are spheres, their
intersection results in a circle.  The sign of the square of the
trivector $L$ immediately encodes whether or not the spheres
intersect.  If $L^2 <0$ the spheres do not intersect, as there are no
null solutions to $L\wdg X =0$ when $L^2 <0$.

As well as new intersection algorithms, the application of geometric
algebra in conformal space provides a highly compact encoding of
reflections.  As a single example, suppose that we wish to reflect the
line $L$ in the plane $P$.  To start with, suppose that the
point of intersection is the origin, with $L$ a straight line in the
$a$ direction, and $P$ the plane defined by the points $b$, $c$ and
the origin.  In this case we have
\begin{equation}
L = a N, \qquad P = (b \wdg c) N.
\end{equation}
The result of reflecting $a$ in $b \wdg c$ is a vector in the $a'$
direction, where
\begin{equation}
a' = (b \wdg c) a (b \wdg c).
\end{equation}
The conformal representation of the line through the origin in the
$a'$ direction is $L'=a'N$.  In terms of their conformal
representations, we have
\begin{equation}
L' = PLP.
\label{Eplnrf}
\end{equation}
So far this is valid at the origin, but conformal invariance ensures
that it holds for all intersection points.  This expression for $L'$
finds the reflected line in space without even having to find the
intersection point.  Furthermore, the line can be curved, and the same
expression reflects a circle in a plane.  We can also consider the
case where the plane itself is curved into a sphere, in which case the
transformation of equation~\rf{Eplnrf} corresponds to an
\textit{inversion} in the sphere.  Inversions are important operations
in geometry, though perhaps of less interest in graphics applications.
To find the reflected line for the spherical case is also
straightforward, as all one needs is the formula for the tangent plane
to the sphere $S$ at the point of intersection.  This is
\begin{equation}
P = (X \dt S) \wdg n
\end{equation}
where $X$ is the point where the line $L$ intersects the sphere.  The
plane $P$ can then be used in equation~\rf{Eplnrf} to find the
reflected line.

\section{Conclusions} 

Conformal geometry provides an efficient framework for studying
Euclidean geometry because the inner product of conformal vectors is
directly related to the Euclidean distance between points.  The true
power of the conformal framework only really emerges when the subject
is formulated in the language of geometric algebra.  The geometric
product unites the outer product, familiar from projective geometry,
with the inner product of conformal vectors.  Each graded subspace
encodes a different geometric primitive, which provides for good data
typing and aids writing robust code.  Furthermore, the algebra
includes multivectors which generate Euclidean transformations.  So
both geometric objects, and the transformations which act on them, are
contained in a single, unified framework.

The remaining step in implementing this algebra in applications is to
construct fast algorithms for multiplying multivectors.  Much as
projective geometry requires a fast engine for multiplying $4 \times
4$ matrices, so conformal geometric algebra requires a fast engine for
multiplying multivectors.  One way to achieve this is to encode each
multivector via its matrix representation.  For the important case of
$\clg(4,1)$ this representation consists of $4 \times 4$ complex
matrices.  But in general this approach is slower than algorithms
designed to take advantage of the unique properties of geometric
algebra.  Such algorithms are under development by a number of groups
around the world.  These will doubtless be of considerable interest to
the graphics community.

\section*{Acknowledgments}

CD would like to thank the organisers of the conference for all their
work during the event, and their patience with this contribution.  CD
is supported by an EPSRC advanced fellowship.

%\bibliography{../articles,../misc}
%\bibliographystyle{unsrt}  

\end{document}